# Raman spectra and Magnetization of all-ferromagnetic superlattices grown on (110) oriented SrTiO$_3$

B. C. Behera[1], A. V. Ravindra[1], P. Padhan[1], and Prellier Wilfrid[2]

[1] *Department of Physics Indian Institute of Technology Madras Chennai – 600036, India*

[2] *Laboratoire CRISMAT, CNRS UMR 6508, ENSICAEN, 6 Bd du Marehal Juin, F-14050 Caen Cedex, FRANCE.*

Abstract

Superlattices consist of two ferromagnets La$_{0.7}$Sr$_{0.3}$MnO$_3$(LSMO) and SrRuO$_3$(SRO) were grown in (110)-orientation on SrTiO$_3$(STO) substrates. The x-ray diffraction and Raman spectra of these superlattices show the presence of in-plane compressive strain and orthorhombic structure of less than 4 u.c. thick LSMO spacer, respectively. Magnetic measurements reveal several features including reduced magnetization, enhanced coercivity, antiferromagnetic coupling, and switching from antiferromagnetic to ferromagnetic coupling with magnetic field orientations. These magnetic properties are explained by the observed orthorhombic structure of spacer LSMO in Raman scattering which occurs due to the modification in the stereochemistry of Mn at the interfaces of SRO and LSMO.

Study on the interface physics of complex oxide of $La_{0.7}Sr_{0.3}MnO_3$ (LSMO)/$SrRuO_3$ (SRO) superlattice structure is important because of several interesting magnetic properties observed in this system [1-6]. One of the most interesting properties of this superlattice structure is the antiferromagnetic coupling at the interfaces of LSMO and SRO even if both components are ferromagnetic. The antiferromagnetic coupling is explained by the interfacial magnetic and electronic disorders[1, 2]. Another fascinating property is that the antiferromagnetic coupling switches to ferromagnetic coupling by changing the orientation of the field from in-plane to out–of–plane direction of the superlattices[2, 5]. In addition, the enhanced magnetization at 10 K is quenched by reversing the stacking order of the constituents SRO and LSMO for the same thickness configuration of the superlattices[2]. This is explained by the competing effect of distortion in the $MnO_6$ and $RuO_6$ octahedra along the out-of-plane direction due to cumulative stress[2]. The LSMO-SRO superlattice also exhibits magnetocaloric effect leading to an improved relative cooling power [3, 4]. In addition inverted hysteresis loops are observed in the field dependent magnetoresistance of LSMO-SRO superlattice[6]. So far, the attention has been mainly on the (001)-oriented LSMO-SRO superlattice, probably due to the simplicity of growth. However, the ferromagnetic properties of thin films and superlattices of perovskite manganites is strongly reduced at the (001)-oriented interface or surface. These phenomena have been attributed to charge redistribution at the interface due to polar discontinuity and orbital ordering induced by strains or broken symmetry [7, 8]. Note that spin states of the double exchange mechanism [9], which explains the ferromagnetism in mixed valence manganite system is sensitive to the bond length, the bond angle of Mn-O-Mn, and the local density of orbital states near the interface. Furthermore, the change of $e_g$-orbital occupation occurs at the surface or interface induced by the surface symmetry breaking and strain [10]. Such double exchange mechanism and orbital occupation at the interface are strongly dependent on the oriented substrate. The thin films of LSMO [11, 12] and SRO [13, 14] have been grown along (001) and (110) orientation and observed enhancement of magnetic properties such as higher Curie temperature and saturation magnetization of (110) oriented thin film compared to the (001) orientation. Such enhancement of magnetic properties of (110) oriented thin film is due to the less crystal deformation, the more compact layer stacking therefore stronger interlayer spin-spin interaction [15], the faster and anisotropic relaxed unit cells, and the quenching of polar discontinuity[16]. We have grown (110) oriented SRO-LSMO superlattices and studied their crystal and electronic structures and magnetic properties. The results are presented in this letter.

Multitarget pulsed laser deposition with a KrF excimer laser ($\lambda$ = 248 nm) has been used for the synthesis of thin films and superlattice structure of SRO and LSMO on the (110)-oriented $SrTiO_3$ (STO) single crystal substrate. The detail of the deposition process for SRO and LSMO on (001) oriented STO has been described previously [1, 2]. The deposition rate for SRO and LSMO are calibrated individually for each laser pulse of energy density ~ 3 $J/cm^2$ and it seems to be almost the same for both ~ 0.73 Å/pulse. A series of superlattice structures with (110)STO/[14-u.c. SRO/ n u.c. LSMO] configurations for n = 2 or 4 were prepared by repeating the bilayer for 15 times. The structural characterizations of the superlattices were performed by using x-ray diffraction. The Raman spectra were recorded on a Jobin-Yvon LabRAM HR800UV spectrometer instrument equipped with highly efficient thermo-electrically cooled charge coupled device (CCD). The spectra were taken at room temperature in the backscattering configuration using 633 nm emission line of a He-Ne laser with lower than 2mW laser power on the sample surface. The magnetic properties measurements were carried out using a superconducting quantum interface device based magnetometer (Quantum Design MPMS-5). The field cooled temperature dependent magnetization measurement is performed in the presence 0.1 tesla external magnetic field along the in-plane and out-of-plane direction of (110) oriented STO substrate. The in-plane and out-of-plane field dependent magnetization curves are measured at 10 K.

Figure 1 shows the *θ-2θ* x-ray diffraction scans of two samples. These samples show up to third order of satellite peaks. Therefore, apart from the substrate (110) STO peak the presence of satellite reflection in the x-ray spectra of these superlattices confirms the long range periodicity of the superlattice structure with good crystallinity. The Laue fringes observed around the 0th order satellite peak in each superlattice indicate that interfaces are smooth with uniform thickness. We have carried out the quantitative refinement of x-ray diffraction profile using DIFFaX program[1, 2] to quantify the number of unit cells presents in these superlattice structures. The simulated profiles for these superlattices around the (110) reflection of STO are also plotted in Fig. 1. The simulated profiles using calibrated thicknesses are in good agreement with the positions of Kiessig fringes and their relative intensity ratio. The out-of-plane lattice parameter "c" calculated from the 0th order satellite peak positions of superlattice with n = 2 and 4 is 2.7816 and 2.7796 Å respectively. The observed out-of-plane lattice parameter of these superlattice is relatively larger compared to the bulk SRO value 2.7789 Å, indicating the presence of in-plane compressive strain due to the lattice mismatch of STO with bottom layer SRO. We have also calculated the change in out-of-plane lattice parameters ("Δc" with respect to the bulk lattice parameter "$c_b$" of bottom layer) to compare the substrate-induced strain on bottom layer SRO. The percentage of change in Δc, with increase in LSMO layer thickness, decreases indicating the relaxation of superlattice.

The crystal symmetry of these superlattices is studied using Raman scattering. In the perfect cubic perovskite structure all lattice sites have inversion symmetry. Therefore, first order Raman scattering is forbidden. However, the LaMnO$_{3.0}$ (undoped parent compound of La$_{0.7}$Sr$_{0.3}$MnO$_3$) and SrRuO$_3$ show orthorhombic structure with $D_{2h}^{16}$ (Pnma) space group[17, 18]. In LaMnO$_{3.0}$ compound, the octahedra have different Mn-O bond lengths, and are rotated around the cubic [010] and [101] axes. It is normally accepted that a cooperative static Jahn-Teller effect at the Mn$^{3+}$ site is responsible for this distorted perovskite structure of manganites[19]. The primitive unit cell has four formula units giving 60 vibrational degrees of freedom. According to the factor group analysis there are 24 Raman modes ($7A_g$ + $5B_{1g}$ + $7B_{2g}$ + $5B_{3g}$) active from the 60 vibrational modes [17, 18]. On the other hand, the LaMnO$_3$ with excess O$_2$ shows rhombohedral crystal structure, space group $D_{3d}^6$ with two formulas per unit cell[20]. This space group is formed by the rotation of MnO$_6$ octahedra about the cubic [111] directions. These rotations could also occur in opposite directions in adjacent unit cells. Thus, the 30 vibrational degrees of freedom per unit cell result at the zone center. However, only A$_{1g}$ and 4E$_g$ modes are Raman active. When LaMnO$_{3.0}$ doped with Sr, the doped compounds are expected to be crystallised with an orthorhombic with $D_{2h}^{16}$ (Pnma) space group for all concentrations similar to the Ca doped LaMnO$_{3.0}$ or Sr doped PrMnO$_{3.0}$[20]. Experimentally, the Sr doped LaMnO$_{3.0}$ exhibit structural phase transition with the doping concentration[20].

Fig. 2 shows the room temperature Raman spectra of (110)STO/[14-u.c. SRO/ n u.c. LSMO]$_{x15}$ superlattice and (110)STO. A weak second-order (two-phonon) features coming from the STO substrate can be seen between 600 and 700 cm$^{-1}$ and as a background in the range 200–500 cm$^{-1}$. In contrast, the superlattice with n = 2 exhibits strong peaks at about 118 and 373 cm$^{-1}$ and several weak peaks at 160, 198, 249, 308, 562 and 620 cm$^{-1}$. However, as the value of "n" increases to 4 the weak peaks become pronounced. The comparison of Raman spectra of superlattices and STO clearly indicates that the Raman modes of the superlattices arise from the constituents SRO and LSMO. The observed Raman lines are further assigned to definite atomic vibrations based on their symmetry, compared to the phonon frequencies predicted by lattice dynamics calculations [17, 18, 20]. Thus, the peaks at 118, 198, 249, 373 and 562 cm$^{-1}$ are all of $A_g$ symmetry, while the peak at 160 cm$^{-1}$ is associated to the $B_{2g}$ symmetry of SRO in the SRO-LSMO superlattice. Indeed, the peaks at 118 and 160 cm$^{-1}$ represent the motion of Sr ions against the RuO$_6$ octahedra, whereas 198 and 249 cm$^{-1}$ represent the rotation/stretching of the RuO$_6$ octahedra. The peak at 373 cm$^{-1}$ assigned to $A_g$ symmetry represents the apical oxygen vibrations that modulate Ru-O-Ru bond angle. Finally, the peak at higher frequency region 562 cm$^{-1}$, with $A_g$ symmetry, represents the in-phase stretching [18]. Nevertheless, the assignment of the LSMO mode in the superlattice is crucial. In these superlattices the thicknesses of the LSMO are indeed significantly smaller than that of the SRO, and thus SRO dominated Raman scattering is expected. The absence of any Raman modes in the frequency range of 400 to 550 cm$^{-1}$ ruled out the possibility of rhombohedral symmetry of LSMO[20, 21]. Considering the orthorhombic structure of LSMO the Raman lines at 198 and 249 cm$^{-1}$ can be assigned to A$_g$ while 308 and 620 cm$^{-1}$ to B$_{2g}$ mode[17]. The A$_g$ and B$_{2g}$, modes can be classified to apical oxygen bending or rotation of the octahedral and to Mn-O bond stretching, respectively[22]. Consequently, the Raman spectra of these superlattices are found to be more consistent with an orthorhombic perovskite structure with space group $D_{2h}^{16}$. Since the LSMO generally stabilizes in the rhombohedral symmetry, the observed orthorhombic symmetry of LSMO provides evidence for possible modification in the stereochemistry of Mn at the interfaces of SRO and LSMO.

Fig. 3 displays the temperature dependent in-plane and out-of-plane magnetization (M(T)) curves recorded for the n = 2 and 4 superlattice in the presence of 0.1 T field. The onset of in-plane magnetization of superlattice with n = 2 arises at around 300 K, which is relatively smaller than the Curie temperature (T$_c$) of bulk LSMO [23]. When the superlattice is cooled down below 300 K the magnetization increases rapidly followed by a gradual increase, leading to a peak at 160 K. On further cooling, the magnetization of the superlattice decreases gradually down to the lowest temperature. The drop in magnetization, at a temperature which we have marked T$_N$, is due to the interfacial magnetic and electronic disorders in the interfacial region [1, 2]. The observed T$_N$ in the in-plane M(T) shows the presence of antiferromagnetic coupling between the SRO and LSMO layers of the superlattices below 160 K, the T$_c$ of SRO[1-4]. Qualitatively, similar M(T) curve with the onset of in-plane magnetization at higher temperature 330 K is observed for the superlattice with n = 4. The higher T$_c$ value compared to that of the superlattice with n = 2 is attributed to the thicker LSMO layer. Though the T$_c$ of the superlattice is similar for both orientations (001) and (110), a remarkable increase (~ 10 K) of the $T_N$ is observed in the case of (110) oriented superlattice compared to that of (001) orientation[1-4]. Furthermore, the M(T) curves of these superlattices diverge significantly for the magnetic field aligned along in-plane and out-of-plane directions. However, as the superlattice with n = 2 is cooled below 330 K the out-of-plane magnetization increases gradually upto 280 K, then very slowly like a plateau down to 160 K followed by a sudden enhancement in magnetization exactly at 160 K, the $T_c$ of SRO. This temperature (160 K) where the SRO layer becomes ferromagnetic is denoted as $T_C^*$. The appearance of $T_C^*$ in M(T) indicates the existence of ferromagnetic coupling at the interfaces of LSMO and SRO along the out-of plane direction of the superlattice[1-4]. Similar features of out-of-plane temperature dependent magnetization curves occur for the superlattice with n = 4. Note that an enhanced $T_C^*$ is observed in the case of (110) oriented superlattice compared to that of (001) orientation.

The field dependent magnetization curves M(H) measured at 10 K of the superlattices with n = 2 and 4 are shown in Fig. 4. Both superlattices exhibit ferromagnetic hysteresis loops when the field is oriented along the in-plane and out-of plane directions. The magnetization increases monotonically with the increase of field even at 4.5 tesla field [unlike in the case of LSMO-SRO superlattice (Ref. 5)]. The observed monotonic increase of high field magnetization of these superlattices suggests the presence of antiferromagnetic coupling with spin pinning/biasing or spin canting[24]. The magnetization at 4.5 tesla is 1.39 and 1.21 $\mu_B$/u.c. for superlattice with n = 2 and 4, respectively. These observed values of magnetization are lower but close to that of the SRO (1.6 $\mu_B$/Ru [25]). However, it is very low compared to that of the magnetization of LSMO (3.34 $\mu_B$/Mn [26]), which again confirms the presence of antiferromagnetic coupling with spin pinning/biasing or spin canting at the interface regions[1,2]. Further, the in-plane coercive field ($H_C$) is found to be 0.1 and 0.17 tesla, whereas the out-of-plane $H_C$ is 0.8 and 1.3 tesla for superlattices with n = 2 and 4, respectively. The observed $H_C$ values of the superlattices are significantly larger than the $H_C$ values of the SRO ($H_C$ ~ 0.07 and ~ 0.3 tesla along the in-plane and out-of-plane respectively[14]) and LSMO ($H_C$ ~ 0.04 and ~ 0.02 tesla along the in-plane and out-of-plane respectively[12]) grown on (110) oriented STO. In addition, the observed $H_C$ values of the (110) oriented SRO-LSMO superlattice are relatively larger than that of the (001) orientation[2]. Such enhancement of $H_C$ is attributed to the observed orthorhombic perovskite structure of LSMO which provides modification in the stereochemistry of Mn at the interfaces of SRO and LSMO as evidenced from the Raman spectra. The orthorhombic unit cell in LSMO is a result of cooperative $MnO_6$ octahedra tilts/rotations induced by the mismatch between La/Sr-O and $\sqrt{2}$(Mn-O) bond lengths. The bond length (i.e. in-plane lattice parameter) of SRO is larger than that of the LSMO. So at the interfaces in order to allow pseudo-cubic in-plane axes to become longer with respect to the out-of-plane axis, octahedral rotations around the out-of-plane direction have to be significantly reduced or absent[27]. Such uniaxial anisotropy is expected to be responsible for the coercivity enhancement.

In conclusion, we have grown (110) oriented superlattices consisting of two metal-like ferromagnets $La_{0.7}Sr_{0.3}MnO_3$ and $SrRuO_3$ on (110) oriented $SrTiO_3$. The x-ray diffraction spectra of these superlattices indicate the presence of in-plane compressive strain due to the lattice mismatch of STO with bottom layer SRO. Whereas the Raman spectra show that the thin film with less than 4 u.c. thick LSMO, when sandwiched between 14 u.c. thick (110) oriented SRO stabilizes in orthorhombic structure. In addition, the SRO-LSMO superlattices when gown on (110) orientation exhibits several features like reduced magnetization, enhanced coercivity, antiferromagnetic coupling, and switching from antiferromagnetic to ferromagnetic coupling with orientation of magnetic field. These magnetic properties of the (110) orientated SRO-LSMO superlattices are explained by the observed orthorhombic structure of spacer LSMO in Raman scattering, which occurs due to the modification in the stereochemistry of Mn at the interfaces of SRO and LSMO. We expect that these results will be useful for an understanding of magnetic coupling when deposited as a multilayer thin film.

**Acknowledgments:**

We greatly acknowledge Indo-French collaboration through the financial support of both the LAFICS and the IFPCAR/CEFIPRA (Project No. 3908-1).

Figure Captions:

Figure 1: The θ-2θ x-ray diffraction profiles and simulated spectra of (a) (110)STO/[14-u.c. SRO/2-u.c. LSMO]$_{x15}$ and (b) (110)STO/[14-u.c. SRO/4-u.c. LSMO]$_{x15}$ superlattices. The (110) Bragg's reflection of STO as well as the satellite peaks of the superlattice are indicated.

Figure 2: Raman spectra of (110) oriented STO substrate and superlattices with n = 2 and 4.

Figure 3: Temperature dependent 0.1 tesla field cooled (a) in-plane magnetization and (b) out-of-plane magnetization of the superlattices with n = 2 and 4. The arrows indicate the $T_C$, $T_C^*$, and $T_N$.

Figure 4: Field dependent magnetization at 10 K with field oriented along the plane and out-of-plane of (a) (110)STO/[14-u.c. SRO/2-u.c. LSMO]$_{x15}$ and (b) (110)STO/[14-u.c. SRO/4-u.c. LSMO]$_{x15}$ superlattices.

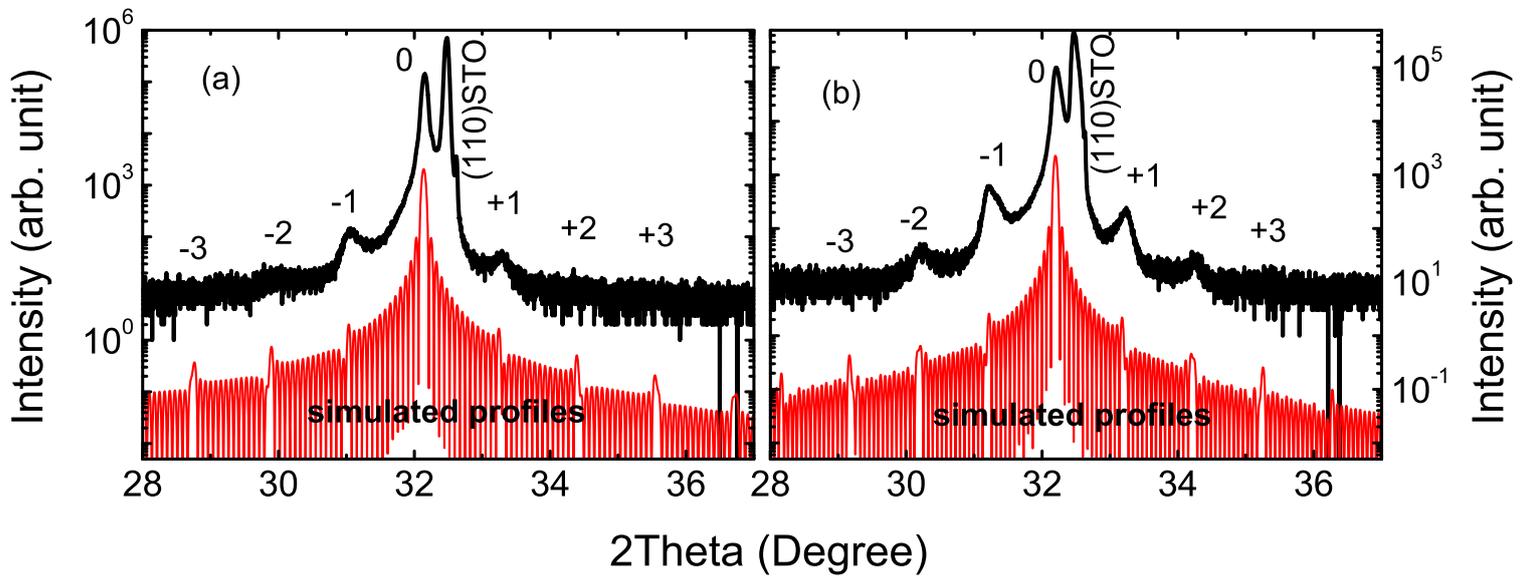

Figure 1: The θ-2θ x-ray diffraction profiles and simulated spectra of (a) (110)STO/[14-u.c. SRO/2-u.c. LSMO]$_{x15}$ and (b) (110)STO/[14-u.c. SRO/4-u.c. LSMO]$_{x15}$ superlattices. The (110) Bragg's reflection of STO as well as the satellite peaks of the superlattice are indicated.

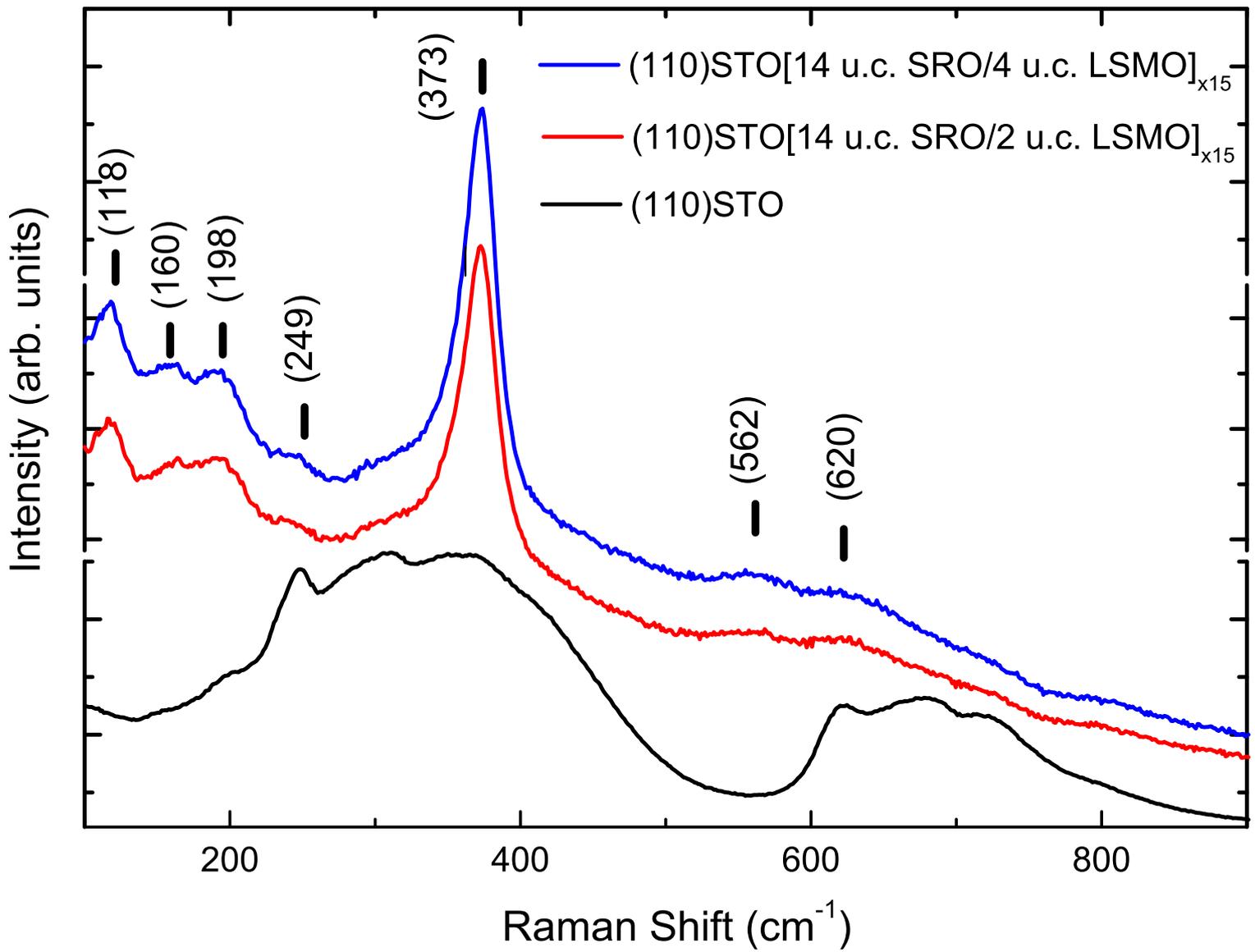

Figure 2: Raman spectra of (110) oriented STO substrate and superlattices with n = 2 and 4.

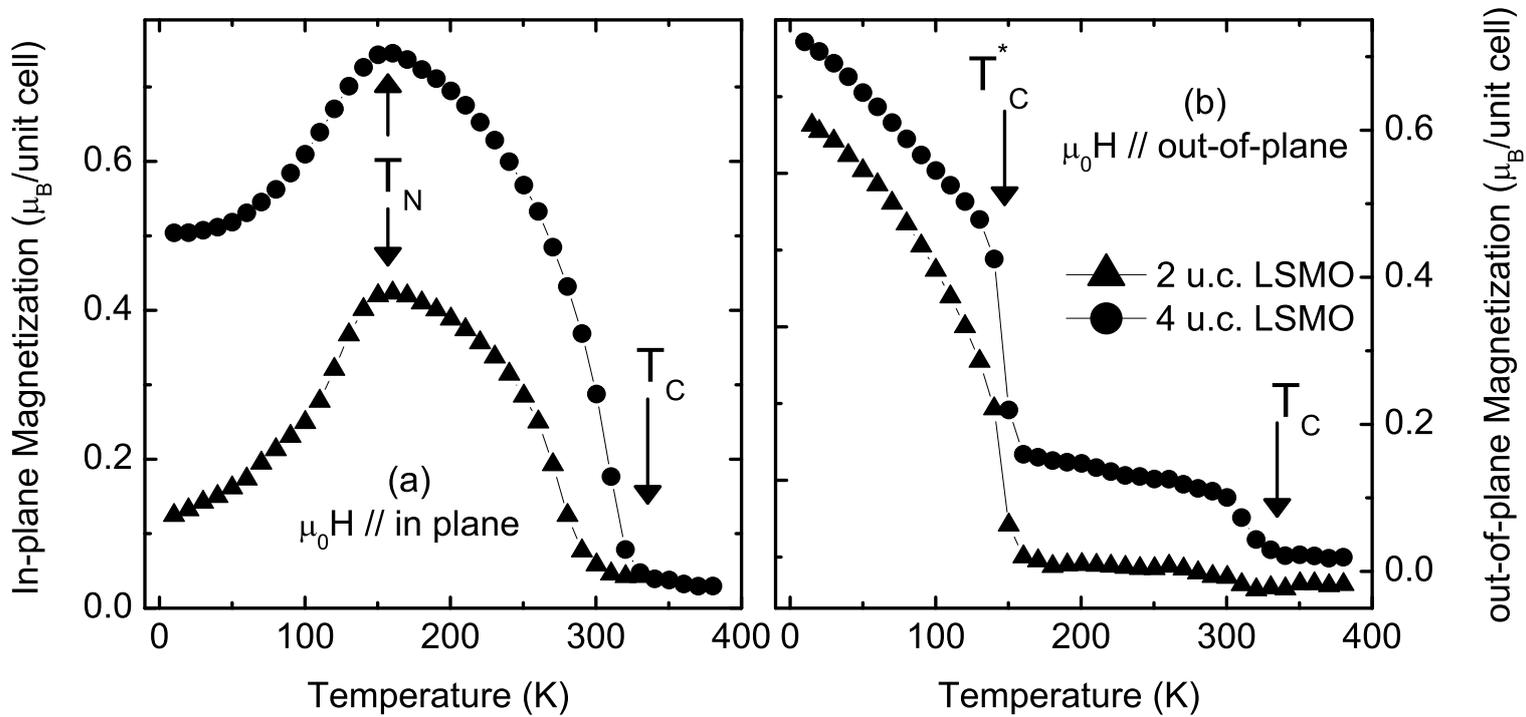

Figure 3: Temperature dependent 0.1 tesla field cooled (a) in-plane magnetization and (b) out-of-plane magnetization of the superlattices with n = 2 and 4. The arrows indicate the $T_C$, $T_C^*$, and $T_N$.

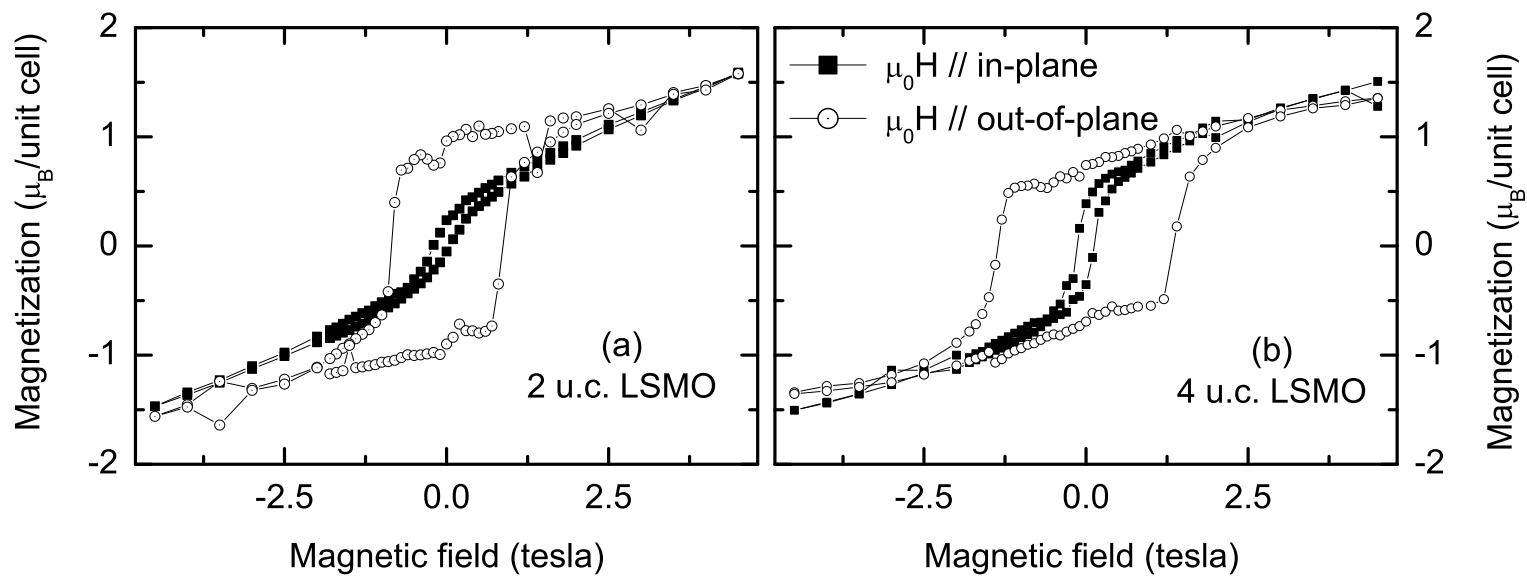

Figure 4: Field dependent magnetization at 10 K with field oriented along the plane and out-of-plane of (a) (110)STO/[14-u.c. SRO/2-u.c. LSMO]$_{x15}$ and (b) (110)STO/[14-u.c. SRO/4-u.c. LSMO]$_{x15}$ superlattices.